\documentstyle[12pt,psfig]{article}
\catcode`\@=11
\def\marginnote#1{}
\newcount\hour
\newcount\minute
\newtoks\amorpm
\hour=\time\divide\hour by60
\minute=\time{\multiply\hour by60 \global\advance\minute by-
\hour}
\edef\standardtime{{\ifnum\hour<12 \global\amorpm={am}%
    \else\global\amorpm={pm}\advance\hour by-12 \fi
    \ifnum\hour=0 \hour=12 \fi
    \number\hour:\ifnum\minute<100\fi\number\minute\the\amorpm}}
\edef\militarytime{\number\hour:\ifnum\minute<100\fi\number\minute}
\def\draftlabel#1{{\@bsphack\if@filesw {\let\thepage\relax
  \xdef\@gtempa{\write\@auxout{\string
    \newlabel{#1}{{\@currentlabel}{\thepage}}}}}\@gtempa
    \if@nobreak \ifvmode\nobreak\fi\fi\fi\@esphack}
     \gdef\@eqnlabel{#1}}
\def\@eqnlabel{}
\def\@vacuum{}
\def\draftmarginnote#1{\marginpar{\raggedright\scriptsize\tt#1}}
\def\draft{\oddsidemargin -.5truein
        \def\@oddfoot{\sl preliminary draft \hfil
        \rm\thepage\hfil\sl\today\quad\militarytime}
        \let\@evenfoot\@oddfoot \overfullrule 3pt
        \let\label=\draftlabel
        \let\marginnote=\draftmarginnote

\def\@eqnnum{(\theequation)\rlap{\kern\marginparsep\tt\@eqnlabel}%
\global\let\@eqnlabel\@vacuum}  }
\def\preprint{\twocolumn\sloppy\flushbottom\parindent 1em
        \leftmargini 2em\leftmarginv .5em\leftmarginvi .5em
        \oddsidemargin -.5in    \evensidemargin -.5in
        \columnsep 15mm \footheight 0pt
        \textwidth 250mmin      \topmargin  -.4in
        \headheight 12pt \topskip .4in
        \textheight 175mm
        \footskip 0pt

\def\@oddhead{\thepage\hfil\addtocounter{page}{1}\thepage}
        \let\@evenhead\@oddhead \def\@oddfoot{} \def\@evenfoot{}
}
\def\titlepage{\@restonecolfalse\if@twocolumn\@restonecoltrue\onecolumn
     \else \newpage \fi \thispagestyle{empty}\c@page\z@
        \def\thefootnote{\fnsymbol{footnote}} }
\def\endtitlepage{\if@restonecol\twocolumn \else  \fi
        \def\thefootnote{\arabic{footnote}}
        \setcounter{footnote}{0}}  %\c@footnote\z@ }
\catcode`@=12
\relax
\def\be{\begin{equation}}
\def\ee{\end{equation}}
\def\bea{\begin{eqnarray}}
\def\eea{\end{eqnarray}}
\def\simlt{\stackrel{<}{{}_\sim}}

\def\NPB#1#2#3{{\it Nucl.~Phys.} {\bf{B#1}} (19#2) #3}
\def\PLB#1#2#3{{\it Phys.~Lett.} {\bf{B#1}} (19#2) #3}
\def\PRD#1#2#3{{\it Phys.~Rev.} {\bf{D#1}} (19#2) #3}
\def\PRL#1#2#3{{\it Phys.~Rev.~Lett.} {\bf{#1}} (19#2) #3}
\def\ZPC#1#2#3{{\it Z.~Phys.} {\bf C#1} (19#2) #3}
\def\PTP#1#2#3{{\it Prog.~Theor.~Phys.} {\bf#1}  (19#2) #3}
\def\MPL#1#2#3{{\it Mod.~Phys.~Lett.} {\bf#1} (19#2) #3}
\def\PR#1#2#3{{\it Phys.~Rep.} {\bf#1} (19#2) #3}
\def\RMP#1#2#3{{\it Rev.~Mod.~Phys.} {\bf#1} (19#2) #3}

\def\mst1{m_{\widetilde{t}_1}}
\def\mst2{m_{\widetilde{t}_2}}
\def\mst12{m_{\widetilde{t}_{1,2}}}

\def\msb1{m_{\widetilde{b}_1}}
\def\msb2{m_{\widetilde{b}_2}}
\def\msb12{m_{\widetilde{b}_{1,2}}}

\def\mtilde2{\widetilde{m}^{2}}

\relax
\begin{document}
\setlength{\baselineskip}{3.0ex}
\begin{titlepage}
\phantom{bla}
%\vspace{3.5cm}
\begin{flushright}
SCIPP 95/48\\
IEM--FT--113/95 \\
%hep--ph/9512359 \\
\end{flushright}
\vskip 0.3in
\begin{center}{\large\bf
New and Strong Constraints on the Parameter Space of the MSSM from
Charge and Color Breaking Directions
\footnote{Work supported in part by
the European Union (contract CHRX-CT92-0004) and
CICYT of Spain
(contract AEN94-0928).}  } \\
\vspace*{6.0ex}
{\large J.A. Casas}
\footnote{On leave of absence from Instituto
de Estructura de la Materia, CSIC, Serrano 123, 28006-Madrid,
Spain.}\\
\vspace*{1.5ex}
{\large\it Santa Cruz Institute for Particle Physics, Univ. of California,
Santa Cruz, CA 95064}\\
\vspace{2cm}
Talk given at the SUSY-95 International Workshop\\
``Supersymmetry and Unification of Fundamental Interactions''\\
Palaiseau, Paris, May 1995
\vspace{2cm}
\end{center}
\vskip.5cm
\begin{center}
{\bf Abstract}
\end{center}
%\begin{quote}
%\vbox{ \baselineskip 14pt

Although the possible existence of dangerous charge and color breaking
(CCB) directions in the MSSM has been known since the early 80's, only
particular directions in the field-space have been considered, thus obtaining
necessary but not sufficient conditions to avoid dangerous CCB minima.
Furthermore, the radiative corrections to the potential were not normally
included in a proper way, often leading to an overestimation of the
restrictive power of the bounds. It turns out that when correctly evaluated,
the ``traditional" CCB bounds are very weak. I give here a brief survey of
recent results on this subject, which represent a complete analysis, showing
that the new CCB bounds are very strong and, in fact, there are extensive
regions in the parameter space that become forbidden. This produces important
bounds, not only on the value of $A$, but also on the values of $B$ and
$M_{1/2}$. The form of strongest one of the new bounds, the so called UFB-3
bound, is explicitely given.
%\end{quote}
%}
%\vskip1.cm

\begin{flushleft}
%CERN-TH/95-xxx\\
November 1995 \\
\end{flushleft}

\end{titlepage}
\newpage
It is well known\cite{Frere,Claudson}, \cite{Drees,Gunion,Komatsu,Gamberini}
that the presence of scalar fields with
color and electric charge in supersymmetric (SUSY) theories induces the
possible existence of dangerous charge and color breaking (CCB)
minima.
However,
a complete study of this crucial issue is still lacking. This is
mainly due to two reasons. First, the enormous complexity of the
scalar potential, $V$, in a SUSY theory, which has motivated that
only analyses examining particular directions in the field--space
have been performed. Second, the radiative
corrections to $V$ have not been normally included in a proper way.
Concerning the first point, the tree-level scalar potential, $V_o$,
in the minimal supersymmetric standard model (MSSM) is given by
$V_o=V_F + V_D + V_{\rm soft}$, where $V_F$ and $V_D$ are the F-- and
D--terms respectively and $V_{\rm soft}$ are the soft breaking terms, i.e.
\bea
\label{Vsoft}
V_{\rm soft}&=&\sum_\alpha m_{\phi_\alpha}^2
|\phi_\alpha|^2\ +\ \sum_{i\equiv generations}\left\{
A_{u_i}\lambda_{u_i}Q_i H_2 u_i + A_{d_i}\lambda_{d_i} Q_i H_1 d_i
\right.
\nonumber \\
&+& \left. A_{e_i}\lambda_{e_i}L_i H_1 {e_i} + {\rm h.c.} \right\}
+ \left( B\mu H_1 H_2 + {\rm h.c.}\right)\;\; ,
\eea
in a standard notation.
$V_o$ is extremely
involved since it  has a large number of independent fields and parameters.
Even assuming universality of the soft breaking terms at
the unification scale, there are five undetermined
parameters: $m$, $M$, $A$, $B$, $\mu$, i.e. the universal
scalar and gaugino masses, the universal
coefficients of the trilinear and bilinear scalar terms, and
the Higgs mixing mass, respectively.
(Notice that
$M$ does not appear explicitely in $V_o$, but it does through the
renormalization group equations (RGEs) of all the remaining parameters.)

As mentioned above, the complexity of $V$ has made that only
particular directions in the field-space have been explored, thus
obtaining necessary but not sufficient CCB conditions to avoid
dangerous CCB minima. By far the most extensively used CCB condition is
the ``traditional" bound, first studied in ref.\cite{Frere,Claudson}.
Namely, given a particular trilinear scalar coupling,
e.g. $\lambda_u A_u Q_u H_2 u$, assuming
equal vacuum expectation values (VEVs) for the three fields involved
in it, i.e. $|Q_u| = |H_2| = |u|$, it turns out that
a very deep CCB minimum appears {\em unless} the famous constraint
\be
\label{frerebound}
|A_u|^2 \leq 3\left( m_{Q_u}^2 + m_{u}^2 + m_2^2\right)
\ee
is satisfied. In the previous equation $m_{Q_u}^2, m_{u}^2, m_2^2$ are the
mass parameters of $Q_u$, $u$, $H_2$, where
$m_2^2$ is the sum of the $H_2$ squared soft mass, $m_{H_2}^2$, plus
$\mu^2$. Further analogous constraints have been derived in the
existing literature \cite{Drees,Gunion,Komatsu,Gamberini}.

Concerning the radiative corrections it should be noted that the usual
CCB bounds (e.g. eq.(\ref{frerebound})) come from the tree-level potential,
$V_o$. However, $V_o$ is strongly dependent on the renormalization scale,
$Q$ and the one-loop radiative corrections to it, namely
$\Delta V_1={\displaystyle\sum_{\alpha}}{\displaystyle\frac{n_\alpha}{64\pi^2}}
M_\alpha^4\left[\log{\displaystyle\frac{M_\alpha^2}{Q^2}}
-\frac{3}{2}\right]$
are crucial to make it stable against variations of $Q$
\cite{Gamberini,CC}. In the previous expression $M_\alpha$ are
the tree-level mass eigenstates, which in general are field-dependent
quantities, so
$\Delta V_1$ is a complicated function of all the scalar fields.
However, a good approximation is to still work just
with $V_o$, but at an appropriate choice for the value of $Q$,
so that $\Delta V_1$ is
small and the predictions of $V_o$ and $V_o+\Delta V_1$ essentially coincide.
This occurs for a value of $Q$ of the order of the most significant $M_\alpha$
mass appearing in $\Delta V_1$, which in turn depends on what is the
direction in the field-space that is
being analyzed. In the usual calculations, however, the CCB bounds
are imposed at any scale between $M_X$ and $M_Z$ and, therefore, their
restrictive power has been overestimated.

In this talk I give a brief survey of the results of our article, ref.
\cite{CCB}, where we have tried to completely classify all the possible
dangerous directions in the MSSM, extracting the corresponding improved
(and hopefully complete) bounds and analyzing numerically their restrictive
power.

It is important to keep in mind that the Higgs part of the potential
must be in such a way that it developes a {\em realistic} minimum at
$|H_1|=v_1$, $|H_2|=v_2$, with
$v_1^2+v_2^2=2M_W^2 / g_2^2$, which
corresponds to the standard vacuum.
This requirement fixes the value of $\mu$ in terms of the
other independent parameters, i.e. $m,M,A,B,\mu$.
Furthermore one has to demand that all the physical
particles have masses compatible with their observed values
(or upper bounds).

There are two types of charge and color breaking constraints:
the ones arising from directions in the field-space along
which the (tree-level) potential can become unbounded from below (UFB),
and those arising from the existence of charge and color
breaking (CCB) minima in the potential deeper than the
realistic minimum. A complete classification of the UFB and CCB
constraints can be obtained from ref. \cite{CCB}. Since there
is no room here to list the precise form of these bounds,
let us mention here their most important characteristics and
what is the most important bound.

Concerning the UFB directions (and corresponding constraints),
there are three of them, labelled as UFB-1, UFB-2, UFB-3
in \cite{CCB}. The relevant scalar fields involved are
$\{H_1,H_2\}$, $\{H_1,H_2, L\}$, $\{H_2, L,{e_L}_j,{e_R}_j\}$ respectively,
where $L$ is a slepton taking the VEV along the $\nu_L$ direction and
${e_L}_j, {e_R}_j$ are selectrons of the $j-$generation. The
UFB-3 bound turns out to be the {\it strongest} one of {\it all}
the UFB and CCB constraints in the parameter space of the MSSM,
so it deserves to be exposed in greater detail.

\begin{description}
\item[UFB-3]

${}^{}$\\
It is possible, by simple analytical minimization, to write the
value of all the relevant fields along the UFB-3 direction in
terms of the $H_2$ one. Then, for any value of $|H_2|<M_X$ satisfying
\be
\label{SU6}
|H_2| > \sqrt{ \frac{\mu^2}{4\lambda_{e_j}^2}
+ \frac{4m_{L}^2}{g'^2+g_2^2}}-\frac{|\mu|}{2\lambda_{e_j}} \ ,
\ee
the value of the potential along the UFB-3 direction is simply given
by
\be
\label{SU8}
V_{\rm UFB-3}=(m_2^2 -\mu^2+ m_{L}^2 )|H_2|^2
+ \frac{|\mu|}{\lambda_{e_j}} ( m_{L_j}^2+m_{e_j}^2+m_{L}^2 ) |H_2|
-\frac{2m_{L}^4}{g'^2+g_2^2} \ ,
\ee
otherwise
\be
\label{SU9}
V_{\rm UFB-3}= (m_2^2 -\mu^2 ) |H_2|^2
+ \frac{|\mu|} {\lambda_{e_j}} ( m_{L_j}^2+m_{e_j}^2 ) |H_2| + \frac{1}{8}
(g'^2+g_2^2)\left[ |H_2|^2+\frac{|\mu|}{\lambda_{e_j}}|H_2|\right]^2 \ .
\ee
In eqs.(\ref{SU8},\ref{SU9}) $\lambda_{e_j}$ is the leptonic Yukawa
coupling of the $j-$generation (see eq.(\ref{Vsoft}). Then, the
UFB-3 condition reads
\be
\label{SU7}
V_{\rm UFB-3}(Q=\hat Q) > V_{\rm real \; min} \ ,
\ee
where $V_{\rm real \; min}=-\frac{1}{8}\left(g^2 + g'^2\right)
\left(v_2^2-v_1^2\right)^2$ is the realistic minimum
and the $\hat Q$ scale is given by
$\hat Q\sim {\rm Max}(g_2 |e|, \lambda_{top} |H_2|,
g_2 |H_2|, g_2 |L|, M_S)$
with
$|e|$=$\sqrt{\frac{|\mu|}{\lambda_{e_j}}|H_2|}$ and
$|L|^2$=$-\frac{4m_{L}^2}{g'^2+g_2^2}$+($|H_2|^2$+$|e|^2$).
Finally $M_S$ is the typical scale of SUSY masses (normally a good
choice for $M_S$ is an average of the stop masses, for more details
see refs.\cite{Gamberini,CC,CCB})
{}From (\ref{SU8}-\ref{SU7}), it is clear that the larger
$\lambda_{e_j}$ the more restrictive
the constraint becomes. Consequently, the optimum choice of
the $e$--type slepton should be the third generation one, i.e.
${e_j}=$ stau.

\end{description}

Let us briefly turn to the CCB constraints in the strict sense, i.e. those
coming from the possible existence of charge and color
breaking (CCB) minima in the potential deeper than the
realistic minimum. We have already mentioned the ``traditional" CCB
constraint \cite{Frere} of eq.(\ref{frerebound}).
Other particular CCB constraints have been explored in the
literature \cite{Drees,Gunion,Komatsu,Langacker}.
In ref.\cite{CCB} it has been performed a complete analysis of
the CCB minima, obtaining a set of ``improved" analytic constraints
that represent the
necessary and sufficient conditions to avoid the dangerous ones.
For certain regions of values of the initial parameters, the CCB constraints
``degenerate" into the above-mentioned UFB constraints since the
minima become unbounded from below directions. In this sense, the
CCB constraints comprise the UFB bounds, so the latter can be
considered as special (but extremely important) limits of the former.

It is not possible to give here an account of the general CCB constraints
obtained in ref.\cite{CCB}, so let us mention their most outstanding
characteristics. First, the most dangerous, i.e. the
deepest, CCB directions in the MSSM potential involve only one particular
trilinear soft term of one generation. Then, for each trilinear soft term
there are three possible (optimized) types of constraints, which in
\cite{CCB} were named CCB-1,2,3. Of course these constraints, which
have an analytical form not very different from the ``traditional"
ones (see eq.(\ref{frerebound})), include the latter and are much
more stronger than them. It is important to recall here that the
CCB bounds must be evaluated at a correct renormalization scale,
$\hat Q$, in order to avoid an overestimation of their restrictive power.
That scale is always of order $\frac{A}{4\lambda}$, where $A$ and
$\lambda$ are, respectively, the coefficient of the trilinear scalar
term and the Yukawa coupling constant associated to the Yukawa coupling
under consideration (a more precise recipe for the value of $\hat Q$
can be found in \cite{CCB}). The numerical analysis shows that the
the ``traditional" CCB bounds when correctly evaluated
turn out to be very weak (see Fig.1a). On the contrary,
the new improved CCB-1,2,3 constraints obtained in ref.\cite{CCB}
(and not explicitely written here) are much more restrictive (see Fig.1b).

Anyway, as mentioned above, the most important restrictions come from
the UFB constraints, in particular from the UFB-3 one, explicitely shown
in eqs.(\ref{SU6}-\ref{SU7}). This is clearly exhibited in Fig.2,
where we summarize all the constraints
plotting also the excluded region due to (conservative)
experimental bounds on SUSY
particle masses. The allowed region left at the end of the day
(white) is quite small.

When the whole MSSM parameter space is scanned it is observed that, as
a general trend, the smaller the value of $m$, the more restrictive the
constraints become. In the limiting case $m=0$ essentially the
{\it whole} parameter space turns out to be excluded. This has obvious
implications, e.g. for no-scale models. Let us also mention that,
contrary to a common believe, the UFB and CCB constraints are
very strong and
put  important bounds not only on the
value of $A$ (soft trilinear parameter), but also on the values of $B$
(soft bilinear parameter) and $M$ (gaugino masses). This is a new and
interesting feature.

\newpage
%\hspace{0.1cm}
%
%\vspace{-0.1cm}

%\newpage
%%%%%%%%%%%%%%%%%%%%%%%%%%%%%%%%%%%%%%%%%%%%%
\begin{figure}[htb]
%\psdraft
\centerline{
\psfig{figure=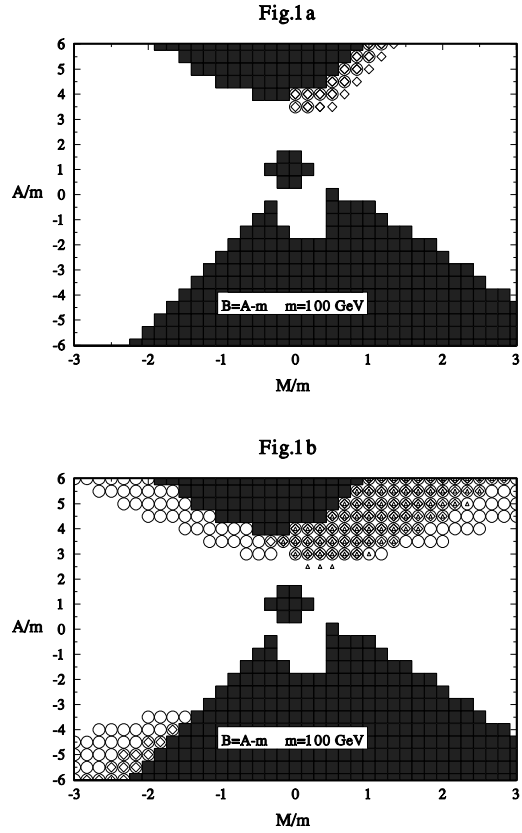,height=11.5cm,angle=180
%,bbllx=9.5cm,bblly=1.cm,bburx=19.cm,bbury=14cm
}}
\caption{{\bf Fig.1:} Excluded regions in the parameter space of the Minimal
Supersymmetric Standard Model, with $B=A-m$, $m=100$ GeV and
$M^{\rm phys}_{\rm top}=174$ GeV. The darked region is excluded because
there is no solution for $\mu$ capable of producing the correct electroweak
breaking. a) The circles and diamonds indicate regions excluded by the
``traditional"
CCB constraints associated with
the $e$ and $d$-type trilinear terms respectively.
b) The same as (a) but using our ``improved" CCB
constraints. The triangles correspond to the $u$-type trilinear terms.}
\end{figure}
%%%%%%%%%%%%%%%%%%%%%%%%%%%%%%%%%%%%%%%%%%%%%%%

\vspace{-3cm}

%\end{document}
%%%%%%%%%%%%%%%%%%%%%%%%%%%%%%%%%%%%%%%%%%%%%
\begin{figure}[htb]
%\psdraft
\centerline{
\psfig{figure=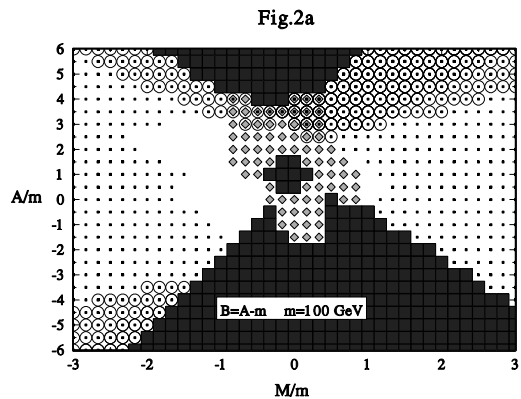,height=11.5cm,angle=180
%,bbllx=9.5cm,bblly=1.cm,bburx=19.cm,bbury=14cm
}}
\vspace{-3.5cm}
\caption{{\bf Fig.2:} Excluded regions in the parameter space of the Minimal
Supersymmetric Standard Model, with $B=A-m$ and
$M^{\rm phys}_{\rm top}=174$ GeV.
The small filled squares indicate regions excluded by our
Unbounded From Below constraints, mainly the UFB-3 one.
The circles indicate regions excluded
by our ``improved" CCB constraints. The filled diamonds
indicate regions excluded by the experimental lower bounds on supersymmetric
particle masses.}
\end{figure}
%%%%%%%%%%%%%%%%%%%%%%%%%%%%%%%%%%%%%%%%%%%%%%%

\newpage

\section*{Acknowledgements}

I thank my collaborators A. Lleyda and C. Mu\~noz for an enjoyable
work in this project

%%%%%%%%%%%%%%%%%%--- References
%%%%%%%%%%%%%%%%%%%%%%%%%%%%%%%%%%%%%%%%%%%%%%%%%%%%%%%
\def\MPL #1 #2 #3 {{\em Mod.~Phys.~Lett.}~{\bf#1}\ (#2) #3 }
\def\NPB #1 #2 #3 {{\em Nucl.~Phys.}~{\bf B#1}\ (#2) #3 }
\def\PLB #1 #2 #3 {{\em Phys.~Lett.}~{\bf B#1}\ (#2) #3 }
\def\PR #1 #2 #3 {{\em Phys.~Rep.}~{\bf#1}\ (#2) #3 }
\def\PRD #1 #2 #3 {{\em Phys.~Rev.}~{\bf D#1}\ (#2) #3 }
\def\PRL #1 #2 #3 {{\em Phys.~Rev.~Lett.}~{\bf#1}\ (#2) #3 }
\def\PTP #1 #2 #3 {{\em Prog.~Theor.~Phys.}~{\bf#1}\ (#2) #3 }
\def\RMP #1 #2 #3 {{\em Rev.~Mod.~Phys.}~{\bf#1}\ (#2) #3 }
\def\ZPC #1 #2 #3 {{\em Z.~Phys.}~{\bf C#1}\ (#2) #3 }

\end{document}